\documentclass[prl,aps,tightenlines,twocolumn,nofootinbib]{revtex4}
\usepackage{graphicx}

\begin{document}

\title{Low scale inflation with large number of e-foldings}

\author{Anupam Mazumdar} 

\affiliation{McGill University, 3600 University Road, Montr\'eal, Qu\'ebec
H3A 2T8, Canada}                        
\vskip15pt
\begin{abstract}
In this paper we illustrate an interesting example of low scale
inflation with an extremely large number of e-foldings.  
This realization can be implemented easily in hybrid inflation model 
where usually inflation ends via phase transition. However this phase 
transition can be so prolong that there is a subsequent epoch of 
slow roll inflation governed by the dynamics of two fields. 
This second bout of inflation can even resolve the $\eta$ problem 
which plagues certain kind of inflationary models. However we also
notice that for extremely low scale inflation it is hard to obtain 
the right amplitude for the scalar density perturbations.
In this regard we invoke alternative mechanisms for generating 
fluctuations. We also describe how to ameliorate the cosmological 
moduli problem in this context.
\end{abstract}
%\pacs{PACS numbers: 11.10Kk, 12.90+b, 98.80Cq}
%\vskip2pc]
\maketitle
%%%%%%%%%%%%%%%%%%%%%%%%%%%%%%%%%%%%%%%%%%%%%%%%%%%%%%%%%%%%%%%%%%%%
\section{Introduction}

Low scale inflation has its virtues and challenges~\cite{Linde}. The 
virtues are low reheating temperature, which can avoid thermal and 
non-thermal gravitino problems~\cite{Ellis,Maroto}, solving moduli problem 
by either invoking thermal inflation~\cite{Lyth} or via TeV scale scale 
inflation~\cite{Randall,garcia}. Among the main challenges it is usually 
hard to realize successful baryogenesis scenario, especially if the reheat 
temperature comes out to be below the electroweak scale, however see 
$Q$-ball baryogenesis~\cite{enqvist98}, and for a review see 
Ref.~\cite{Reps}.

Usually low scale inflationary models do not give plenty of 
inflation compared to a simple chaotic type models, 
partly because the vevs of the fields involved can be 
less than ${\cal O}(M_{p})$, where $M_{p}=2.4\times 10^{19}$~GeV. 
Even hybrid~\cite{Linde1}, or natural inflation type models~\cite{new} 
give limited inflation, e.g. ${\cal N}_{e} \sim 100$, or, even 
less~\cite{angular}. Very Large number of e-foldings is not 
mandatory for structure formation, because the required number of 
e-foldings can be at most $60$,  but in any case it is an interesting
academic curiosity that how large e-foldings of inflation can we 
obtain for low scale inflaton.

In this paper we will illustrate an example based on a simple 
hybrid inflation model~\cite{Linde1}, where it is quite possible 
to tweak the parameters which can give an extremely large e-foldings 
of inflation. Usually in hybrid model inflation ends when the phase 
transition occurs and the transition can be either first order or 
second order~\cite{Copeland}. Particularly in the case of a second 
order phase transition it is often believed that it happens quite 
fast, on the contrary to the common dogma here we give an example 
where this phase transition occurs extremely slowly which can  
drive a second bout of inflation.

One of the most interesting outcome of our scenario is that we can get
enough e-foldings of inflation even if we had a Hubble induced 
mass correction to the inflaton~\cite{softmass}. Imagining that we 
did not have any significant inflation before the phase transition
due to the break down of a slow roll inflation, nevertheless, we can 
still generate sufficiently large e-foldings during the slow phase 
transition. This is the most interesting result of our paper, 
which could provide a resolution to many inflationary
models, such as in $N=1$ supergravity where one commonly obtains such
correction~\cite{Copeland,dterm}. In this paper we also illustrate that 
inspite of having large mass correction to the inflaton it is possible 
to generate almost scale invariant density perturbations, which is a 
compelling result. We will also address the issue of cosmological moduli 
problem in this paper.

%%%%%%%%%%%%%%%%%%%%%%%%%%%%%%%%%%%%%%%%%%%%%%%%%%%%%%%%%%%%%%%%%%%%%

\section{The model and its dynamics}

The dynamics of hybrid inflation models where there is a 
second period of inflation have previously been studied in 
Refs~\cite{Randall} and \cite{garcia}, with particular emphasis on
the production of large density perturbations, which may lead to 
the over-production of primordial black holes. The importance of a
second bout of inflation was rediscovered in the context of inflationary 
models with large extra dimensions~\cite{Green}, see also~\cite{Abdel1}. 
Now we will discuss briefly the dynamics of a second bout of inflation 
and describe its rich applications.

For the purpose of illustration let us consider the following
form of the potential
\begin{eqnarray}
\label{eqn2}
V(\phi,\Phi)=\lambda^2\left(\left(\frac{M_{\ast}}{M_{p}}\right)
\frac{\Phi^2}{2}-\left(\frac{M_{p}}{M_{\ast}}\right)\Phi_{0}^2\right)^2
+\nonumber \\
g^2\left(\frac{M_{\ast}}{M_{p}}\right)^2\phi^2\Phi^2+
\frac{1}{2}m_{\phi}^2\phi^2\,.
\end{eqnarray}
where we have assumed $\lambda,~g \sim {\cal O}(1)$. 
Here $M_{\ast}$ is an intermediate scale. The global 
minimum of the potential is
\begin{equation}
\label{min}
\Phi =\sqrt{2}\left(\frac{M_{p}}{M_{\ast}}\right)\Phi_{0}\,,~~~~\phi=0\,.
\end{equation}
If $M_{\ast}=\Phi_{0}$, then the natural global minimum is 
$\Phi=\sqrt{2}M_{p}$, and $\phi=0$ as before. Naturally our 
couplings are small and for such small couplings we will show that 
phase transition is indeed very slow.  Furthermore if 
$\Phi_{0}\gg m_{\phi}$ and for $\Phi\ll (M_{\rm P}\Phi_{0}/M_{\ast})$ 
the false vacuum term in the potential dominates and the Hubble 
parameter remains constant with 
\begin{equation}
H=\sqrt{\frac{8\pi}{3}}\lambda\frac{\Phi_{0}^2}{M_{\ast}}\,.
\end{equation}
During the first phase of inflation when $\phi>\phi_{c},~\Phi=0$, the
number of e-foldings of inflation is limited by
\begin{equation}
\frac{\phi}{\phi_{c}}=e^{\frac{n-1}{2}{\cal N}_{e}}\,,
\end{equation}
where $n$ is the spectral index, and it is given by
\begin{equation}
\label{tilt}
n-1=2\eta=\frac{2m_{\phi}^2}{3H^2}\,.
\end{equation}
Note that if $m_{\phi}\sim {\cal O}(H)$, the tilt in the spectral index will
be significantly large. This is exactly what happens in the context of a
Hubble induced mass correction to the inflaton in $F$-term supergravity
inflation~\cite{Copeland,dterm}.
In the above expression $\phi_{c}$ is the critical
vev which is determined by
\begin{equation}
\phi_{c}\equiv \frac{\lambda\Phi_{0}M_{p}}{g M_{\ast}}\,.
\end{equation}
This is the point where usually the slow roll inflation comes to an 
end with a phase transition. For the time being let us concentrate upon
the dynamics around this vev.

Note that the slope of the potential in the $\phi$ direction 
can be written as
\begin{equation}
\label{deriv1}
\frac{{\rm d}V}{{\rm d} \phi} = \left[2g^2\left(\frac{M_{*}}{M_{\rm P}} 
\right)^{2} \Phi^{2}+m_{\phi}^{2}\right] \phi \,.
\end{equation}
When the second term in the right hand side of the above equation dominates  
then the motion of $\phi$ evolves away from $\phi_{\rm c}$.
By solving the equation of motion for $\phi$, we obtain
\begin{equation}
\label{phi2}
\phi=\phi_{c} \exp{ \left[ - \frac{1}{\sqrt{24 \pi}} \frac{1}{\lambda}
        \left( \frac{m_{\phi}}{\Phi_{0}} \right)^{2} (M_{\ast} \hat{t})
        \right]} \,,
\end{equation}
where we have now taken $\hat{t}=0$ when $\phi=\phi_{\rm c}$.
For small vevs the quantum fluctuations in the $\Phi$ field become important.
This happens when
\begin{equation}
\phi^2 < \phi_{{\rm c}}^2 +\frac{4\pi}{3}\left(\frac{\lambda\Phi_{0}^2
         M_{\rm P}}{M_{\ast}}\right)^2 \,,
\end{equation}
then the Fokker-Planck equation~\cite{fp} is usually employed to study the
dynamics of the $\Phi$ field~\cite{Randall,garcia,Green}. It is 
possible to study the random walk of $\Phi$ field by assuming that 
the field has a delta-function distribution at some initial time,
say when $\phi\gg\phi_{{\rm c}}$. The average quantum diffusion per 
Hubble volume per Hubble time is $\approx H/2\pi$.  It was found 
in Ref.~\cite{garcia} that the typical value of the $\Phi$ field 
when $\phi=\phi_{\rm c}$ is $\bar{\Phi}\sim {\cal O}(10) H/2\pi $,
which has also been numerically verified in Ref.~\cite{Green}.

In the $\Phi$ direction the slope of the potential is given by
\begin{equation}
\label{deriv2}
\frac{{\rm d}V}{{\rm d} \Phi} = \left[\lambda^2\Phi^{2} + 2g^2(\phi^{2} 
-\phi_{c}^{2} ) \right] \Phi \left( \frac{M_{*}}{M_{\rm P}} \right)^{2} \,.
\end{equation}
If $\bar{\Phi}$ is sufficiently large then there is a small period
where the first term dominates and $\Phi(\hat{t})\sim \bar{\Phi}$. As
$\phi$ evolves away from $\phi_{{\rm c}}$, however, the second term
soon comes to dominate, so that for small $\hat{t}$, using the first
order expansion of Eq.~(\ref{phi2})
\begin{equation}
\phi-\phi_{c} \sim - \frac{1}{\sqrt{24 \pi}} \frac{1}{g} \left( 
\frac{m_{\phi}^2}{\Phi_{0} }\right)M_{\rm p} t \,,
\end{equation}
the $\Phi$ field grows exponentially
\begin{equation}
\label{n2}
\Phi=\bar{\Phi}\,\exp{ \left[ \frac{1}{12 \pi}\left( \frac{m_{\phi}}{\Phi_{0}} 
\right)^{2} \left( M_{\ast} t \right)^{2} \right]}  \,,
\end{equation}
where we have neglected the initial period where $\Phi \sim \bar{\Phi}$
because its duration is negligible compared with that of the subsequent
exponential growth.  During this period $\Phi$ field grows exponentially 
and $\phi$ moves slowly away from $\phi_{{\rm c}}$. However, once
\begin{equation}
\label{end}
2g^2\left( \frac{M_{*}}{M_{\rm P}}\right)^{2}\Phi^{2}\sim m_{\phi}^{2}\,,
\end{equation}
the first term in Eq.~(\ref{deriv1}), which is growing
exponentially, comes to dominate the evolution of the $\phi$ field then
$\phi$ evolves rapidly away from $\phi_{\rm c}$. At
this point $\Phi \sim (\phi_{{\rm c}}-\phi)$ so that ${\rm d}V/{\rm d}\Phi
\sim {\rm d}V/{\rm d}\phi$ and both $\Phi$ and $(\phi_{{\rm c}}-\phi)$
grow rapidly, and inflation comes to an end shortly afterwards, with
both fields subsequently oscillating about the global minimum of the
potential. This is the end of the slow phase transition and 
also the era of the end of inflation. The duration of the second phase 
of inflation, $\Delta t$, lasts~\cite{Green}
\begin{equation}
\label{time}
\Delta t\approx \frac{\sqrt{6 \pi}  \Phi_{0}}{m_{\phi} M_{\ast}} 
\left\{\ln\left[ \frac{1}{2g^2}\left(\frac{M_P}{M_{\ast}}\right)^2\left
(\frac{m_{\phi}}{\bar \Phi}\right)^2\right]\right\}^{1/2}\,.
\end{equation}
Since the Hubble parameter remains constant until very close to the end
of the second phase of inflation, the estimate of the total number of
e-foldings can be given by~\cite{Green}
\begin{eqnarray}
\label{ne2}
{\cal N}_{\rm e} & \approx & H \Delta t = 4 \pi\lambda
\frac{\Phi_{0}^3}{M_{\ast}^2 m_{\phi}} \nonumber \\
&&\times \left\{\ln\left[
2\sqrt{\pi}\left(\frac{\Phi_0}{M_{\ast}}\right)^2\left(\frac{M_{\rm
p}}{\bar \Phi}\right)\right]\right\}^{1/2}\,.
\end{eqnarray}
Let us consider some examples, for $\Phi_0\sim M_{\ast}\sim 10^{18}$~GeV,
$m_{\phi}\sim 1$~MeV and $\lambda\sim g\sim {\cal O}(1)$, we obtain 
${\cal N}_{e}\sim 10^{28}$, an extremely large number of e-foldings of 
inflation. This reiterates the point that for $\Phi_0 \sim M_{\ast}$ 
the phase transition is extremely slow. However note that the Hubble
expansion is fairly large, $H\sim 10^{17}$~GeV. Now suppose we consider 
$\Phi_{0}\sim M_{\ast}\sim 10^{6}$~GeV and $m_{\phi}\sim 10$~GeV,
for which $H\sim 10^{6}$~GeV, we still get ${\cal N}_{e}\sim 10^{6}$, still
quite large number of e-foldings of inflation. If we would like to stretch 
our parameters to $m_{\phi}\sim 10^{-3}$~eV and 
$\Phi_{0}\sim M_{\ast}\sim 10^{3}$~GeV, we obtain ${\cal N}_{e}\sim 10^{16}$
e-foldings of inflation. Now we turn our attention to the virtues of the
slow phase transition.

%%%%%%%%%%%%%%%%%%%%%%%%%%%%%%%%%%%%%%%%%%%%%%%%%%%%%%%%%%%%%%%

\section{A possible solution to the $\eta\sim {\cal O}(1)$ problem}

It is quite tempting to push this limit to solve the $\eta$ problem, 
which plagues $F$-term supersymmetric inflationary models and other
models. In supergravity, the idea is that even for a minimal K\"ahler 
potential, the inflaton mass obtains a Hubble induced correction during 
and after inflation, e.g. $m_{\phi}\sim H$, provided the source for the 
inflaton potential belongs to the chiral sector. For $D$-term inflation it 
is possible to avoid this potential problem, see~\cite{dterm}.
An another example for Hubble induced mass correction is if the
inflaton, $\phi$, couples non-minimally to the gravity. The coupling 
$(\xi/2)R\phi^2$ gives rise to the inflaton mass correction, $12\xi H^2$
during inflation, without altering the rate of expansion of the Universe.
Whatsoever be the main cause for obtaining large mass correction, the
main result is that we immediately obtain a large tilt in the 
spectral index, see Eq.~(\ref{tilt}). The inflaton with a Hubble induced 
mass simply rolls down to its minimum within one Hubble time, therefore, 
ending slow roll inflation in usual sense.

Let us now consider our potential, Eq.~(\ref{eqn2}), with 
$m_{\phi}\sim {\cal O}(H)$.
Note that in Eq.~(\ref{ne2}), if we assume $m_{\phi}\sim {\cal O}(1)H$, 
then we obtain ${\cal N}_{e}\sim {\cal O}(10)(\Phi_{0}/M_{\ast})$. For 
instance if we select $\Phi_{0}\sim 10M_{\ast}$, we obtain a considerable 
e-foldings of inflation from a slow phase transition,  
${\cal N}_{e}\sim {\cal O}(100)$. However we immediately note that for 
the above parameters our approximations leading to Eq.~(\ref{ne2}) breaks 
down. With the above parameters phase transition occurs actually fast 
and therefore the problem remains.

Nevertheless it is possible to ameliorate this situation 
if we allow mismatching couplings, for example if we only alter 
\begin{equation}
g= \left(\frac{M_{p}}{M_{\ast}}\right)\,,
\end{equation}
in which case the coupling between the fields simply reads; $\phi^2\Phi^2$
in Eq.~(\ref{eqn2}), and similarly all other equations are modified, 
importantly Eq.~(\ref{end}), which now reads $\Phi^2\sim m^2_{\phi}/2$.
If we suppose $m_{\phi}\sim {\cal O}(H)$, then we obtain a vev of 
$\Phi\sim (\lambda/2)(\Phi_{0}^2/M_{\ast})$, when roughly inflation ends.
Note that this vev is smaller than the global minimum for $\Phi$, 
see Eq.~(\ref{min}), as long as $\Phi_{0}<(2\sqrt{2}/\lambda)M_{p}$.

With the above choice of parameter, $g$, we find an equivalent expression for 
the number of e-foldings to be
\begin{equation}
{\cal N}_{e}\approx 4\pi\frac{\Phi_{0}}{M_{\ast}}\left\{\ln\left[ 
\frac{1}{2}\left(\frac{m_{\phi}}{\bar \Phi}\right)^2\right]\right\}^{1/2}\,.
\end{equation}
We note that we can obtain $\approx 60$ e-foldings of inflation if we just
assume $\Phi_{0}\sim 5\times M_{\ast}$. The number of e-foldings can 
easily increase if $M_{p}\simeq\Phi_{0}\gg M_{\ast}$. We have ensured
numerically that for a wide ranging parameters $m_{\phi}/\bar\Phi>\sqrt{2}$.

The conclusion is robust which reiterates that indeed the phase 
transition is sufficiently slow to have enough e-foldings of 
inflation even if there is a Hubble induced mass correction to the 
inflaton.

%%%%%%%%%%%%%%%%%%%%%%%%%%%%%%%%%%%%%%%%%%%%%%%%%%%%%%%%%

\section{What happens to the cosmological moduli problem?}

The moduli problem occurs with light scalar particles having mass within 
a range $m_{3/2}\leq 1$~TeV, oscillating with an initial vev larger 
than or close to $M_{p}$, with a Planck suppressed interactions to the 
Standard Model degrees of freedom. The small decay rate allows the 
field to oscillate many times before it decays, while oscillating 
the equation of state mimics that of a dust and soon the coherent 
energy density stored in the moduli dominates the Universe, i.e. 
$\rho \sim H^2M_{p}^2$. This is known as the moduli problem. If 
the moduli mass is greater than $10$~TeV then there is no moduli 
problem because they decay before BBN, causing no harm
to the predictions of the hot big bang model. For a supersymmetric
moduli if there is a low scale inflation, e.g. $H\sim {\cal O}(\rm TeV)$,
then it is possible to dilute the moduli abundance, this was the suggestion 
made in  Refs.~\cite{Randall,Lyth}. 

In our case it is quite possible to tune the parameters to obtain 
TeV scale inflation from the slow phase transition. For an example, 
$M_{\ast}\sim \Phi_{0}\sim {\cal O}({\rm TeV})$ and $m_{\phi}\sim 1$~GeV
can give rise to $H\sim {\rm TeV}$ and ${\cal N}_{e}\sim 100$.   
However in models where gravity mediated supersymmetry breaking is 
allowed the natural mass scale comes out to be the electroweak scale. 
In such a case it is rather difficult to tune parameters 
$M_{\ast},~\Phi_{0}$ to have extremely slow phase transition with 
a large number of e-foldings.

Nevertheless it is possible to have intermediate scale inflation
which helps diluting the moduli by decreasing their initial 
amplitude from ${\cal O}(M_{p})$, but it actually does not solve 
problem, because the moduli vev changes during and after inflation.
However if there is an enhanced symmetry, which allows local minimum 
during inflation and the true minimum of the moduli coinciding, 
then the moduli field can be dynamically relaxed to its minimum
during inflation~\cite{Thomas}. In this paper we will consider moduli
induced isocurvature density perturbations later on.

%%%%%%%%%%%%%%%%%%%%%%%%%%%%%%%%%%%%%%%%%%%%%%%%%%%%%%%%%%%%%%%%%%%%%%
\section{Density perturbations}

The amplitude for the density perturbations towards the end of the 
first period of inflation, when $\phi\geq \phi_{\rm c}$ and $\Phi\sim 0$, 
can be calculated easily
\begin{equation}
\label{imp1}
\delta_{\rm H}\sim 
8.2\lambda^2g\frac{\Phi_0^5}{M_{\ast}^2m_{\phi}^2 M_{\rm p}}\,.
\end{equation}
However as we have shown that number of foldings could be fairly large if the
phase transition is extremely slow, in which case the above estimation 
will not hold any significance. One has to invoke the dynamics of both the
fields, we therefore need to employ the formula for multiple
slow-rolling scalar fields~\cite{msr}:
\begin{equation}
\label{estim1}
\delta_{\rm H}^2  = \frac{1}{75 \pi^2} \left( \frac{8 \pi}{M_{\rm p}^2} 
                  \right)^3  V^3 \left[ \left( \frac{{\rm d}V}{{\rm d} \phi}
                   \right)^2 +  \left( \frac{{\rm d}V}{{\rm d} \Phi}
                   \right)^2 \right]^{-1} \,.
\end{equation}
The scales that we are interested in leave the Hubble radius very
close to the end of the second period of inflation when both fields 
are evolving rapidly. However note that after the beginning of the 
phase transition, but far from the end of inflation, 
${\rm d}V/{\rm d} \phi \approx m_{\phi}^2 \phi \gg {\rm d}V/{\rm d}\Phi$ 
so that $\delta_{\rm H}$ has the same value as prior to the phase 
transition, given by Eq.~(\ref{imp1}). The fields begin evolving more 
rapidly during the last couple of e-foldings or so with 
${\rm d}V/{\rm d}\Phi$ and ${\rm d}V/{\rm d}\phi$, which are of the same 
order of magnitude, increasing significantly, so that $\delta_{\rm H}$ 
decreases. 

We can make an order of magnitude estimate of $\delta_{\rm H}$ 
at the end of inflation, $\delta_{\rm H}(\epsilon=1)$, by pretending that 
only one of the fields is dynamically important and utilizing the 
single-field expression for $\delta_{\rm H}$ in terms of the first 
slow-roll parameter $\epsilon_{\phi} \equiv(M_P^2/ 16 \pi) (V'/V)^2$:
\begin{equation}
\delta_{\rm H} = \frac{32}{75 M_{\rm P}^4} \frac{V}{\epsilon} \,.
\end{equation}
Inflation ends when $\epsilon=1$, so that
\begin{equation}
\label{delend}
\delta_{\rm H}(\epsilon=1) < \frac{\lambda \Phi_0^2}{2 M_{\ast} M_{\rm P}}
\,.
\end{equation}
We can obtain an amplitude $\delta_{\rm H}\sim 10^{-5}$ for density 
perturbations if we assume $M_{\ast}\sim \Phi_{0}$ and 
$M_{\ast}\sim 10^{15}$~GeV, for $\lambda\sim 1$. Nevertheless we 
warn that the above estimation is an upper bound only.
In fact we can imagine a situation where we have large number of e-foldings
with small parameters, $M_{\ast},~\Phi_{0}$, such that we do not 
obtain the right amplitude, e.g. $\Phi_{0}\sim M_{\ast}\sim 10^{6}$~GeV 
and $m_{\phi}\sim 10$~GeV, we obtain $\delta_{H}\leq 10^{-12}$.
In such cases we will need to invoke alternative mechanisms for 
generating adiabatic density perturbations.

%%%%%%%%%%%%%%%%%%%%%%%%%%%%%%%%%%%%%%%%%%%%%%%%%%%% 

\section{Alternative mechanisms for generating adiabatic density perturbations}

\subsection{Curvaton Scenario}

It is possible to obtain adiabatic perturbations on comoving 
scales larger than the size of the horizon by some other light degrees of
freedom which does not necessarily belong to the inflaton sector.
This is the main motive behind the curvaton scenario
\cite{Sloth,Lyth1,Moroi,Mazumdar,mcdonald,Lyth2}. Especially the last
reference, \cite{Lyth2}, describes the curvaton scenario in a low scale
inflation models. Let us briefly recall here the outline of 
the curvaton scenario. The perturbations in the light degrees 
of freedom can be regarded as isocurvature in nature. On large scales 
isocurvature perturbations feed the adiabatic fluctuations. The isocurvature 
perturbations can be converted into the adiabatic ones when the light
degree of freedom decays into SM radiation. Such a conversion takes 
place when the contribution of the curvaton energy density $\rho$ to 
the total energy density in the universe grows, i.e., with the increase of
\begin{equation}
r =\frac{3 \rho}{4 \rho_\gamma + 3 \rho}\,.
\label{r}
\end{equation}
Here on $\rho_\gamma$ is the energy in the radiation bath from inflaton
decay. Non-Gaussianity of the produced adiabatic perturbations
requires the curvaton to contribute more than $1\%$ to the energy
density of the universe at the time of decay, that is 
$r_{\rm dec} > 0.01$~\cite{Lyth1,WMAP}.

The curvaton potential is flat during inflation leading to a large VEV
of the curvaton.  The field amplitude remains fixed until the flaton
mass becomes of the order of the Hubble constant, and the field starts
oscillating in the potential.  During this period of oscillations the
curvaton energy density red shifts as non-relativistic matter
$\rho_\phi \propto a^{-3}$, whereas the energy density in the
radiation bath from inflaton decay red shifts as 
$\rho_\gamma \propto a^{-4}$.  Therefore $r \propto a$ grows, 
and isocurvature perturbations are transformed into the adiabatic ones.  
This conversion ends when the curvaton comes to dominate the energy
density, or if that never happens, when it decays.

We can imagine that moduli to be our curvaton. For simplicity 
let us suppose that there is no moduli problem, e.g. 
minimum of the moduli during and after inflation is the same, 
otherwise moduli could be heavier than $10$~TeV. 
In these cases we can imagine that the dynamics of the moduli 
field can generate isocurvature density perturbations during 
inflation~\cite{Moroi}. Depending on the mass of the moduli and 
the scale of inflation, the moduli can roll down slowly or faster 
during inflation. The perturbed equation of motion for the moduli 
field in a long wavelength regime, during inflation, is given by
\begin{equation}
\ddot\delta{\cal M}_{k}+3H\dot\delta{\cal M}_{k}+m^2_{{\cal M}}
\delta{\cal M}_{k}=0\,. 
\end{equation}
The perturbations in the moduli is Gaussian with a spectrum 
${\cal P}^{1/2}_{\delta\cal M}\sim H_{\ast}/2\pi$, where $H_{\ast}=k/a$
denotes the epoch of horizon exit. The spectral tilt in the spectrum is
given by
\begin{equation}
n_{{\cal M}}=2\frac{\dot H}{H^2}+\frac{2}{3}\frac{m^2_{\cal M}}{H^2}\,.
\end{equation}
Note that the spectral tilt depends on the moduli mass. If it were of
order $H$, it would generate a wrong tilt in the power spectrum. For
$m_{{\cal M}}< H$, it is possible to have a flat spectrum.

When $m_{{\cal M}}\sim H(t)$, the moduli starts oscillating. During 
this period the power spectrum can be evaluated by averaging over many
oscillations which generates fluctuations which is $\chi^2$ in nature.
The amplitude for the density perturbations is given by~\cite{Lyth1,Moroi}
\begin{equation}
\delta_{\rm H}\sim r\frac{2}{5}\frac{H_{\ast}}{\pi{\cal M}}\,.
\end{equation}
Now there is a more freedom to accommodate the right amplitude for the 
density perturbations. Nevertheless if $H_{\ast}$ is sufficiently low,
for  instance $H\sim 10^{5}$~GeV, then in order to have the right amplitude
the vev of moduli should be smaller compared to the Planck scale, e.g.
${\cal M}< M_{p}$. This can be easily realized inspite of the fact that 
$m_{\cal M}<<H$, because the number of e-foldings can be fairly large
and during slow rolling the moduli vev slides by 
${\cal M}(t)\sim M_{p}e^{-(2m_{\cal M}^2~t/3H)}$.

Instead of moduli we can also involve the dynamics of 
MSSM flat directions~\cite{Mazumdar}. However in which case one has 
to be careful with the flat direction potentials. Usually MSSM flat 
directions are gauge invariant monomials which obtains supersymmetry 
breaking contributions including non-renormalizable superpotential 
corrections. It was found that only those MSSM flat directions which 
are lifted by $n=7,~9$, non-renormalizable superpotential terms 
are the successful candidates~\cite{Mazumdar}.

\vskip5pt
%%%%%%%%%%%%%%%%%%%%%%%%%%%%%%%%%%%%%%%%%%%%%%%%%%%%%%%%%%%%%%%%%%%%

\subsection{Fluctuating Inflaton Mass and Coupling}

In our case when inflation ends after the phase transition then both 
$\Phi,~\phi$ start oscillating in their respective minimum. However note
that the inflaton, $\phi$, field obtains an effective mass term by virtue 
of the coupling with $\Phi$ field and vice versa. The frequency of 
oscillations for both the fields are determined by their effective 
masses at their global minimum
\begin{equation}
\bar m_{\phi}=\sqrt{2}g\Phi_{0}\,,~~~~~\bar m_{\Phi}=\sqrt{2}\lambda\Phi_{0}\,.
\end{equation}
In fact we can imagine a situation where $g>\lambda$, in which case the
inflaton oscillates with a larger frequency compared to the $\Phi$ field.
Therefore the inflaton could decay into the Standard Model fermions at a 
faster rate compared to the $\Phi$ field (such couplings are certainly
required for a successful reheating, for example see~\cite{jaikumar}). 
We can then assume that $\Phi$
field is still rolling down the potential while the $\phi$ had already 
started oscillating and decaying. The effective inflaton mass due to 
the finite coupling obtains a vev dependent contribution
\begin{equation}
m^2_{eff,~\phi}= 2\left(g\frac{M_{\ast}}{M_{p}}\right)^2\Phi^2+
m^2_{\phi}\,.
\end{equation}
As long as $\Phi\geq (m_{\phi}/2g)(M_{p}/M_{\ast})$, the fluctuations in
the decay rate of the inflaton can be written as
\begin{equation}
\frac{\delta \Gamma}{\Gamma}\sim \frac{\delta\Phi}{\Phi}\,.
\end{equation}
Further imagining that if $\Phi$ field has a Gaussian fluctuation then this
will be imprinted in the decay rate of the inflaton as well, e.g. 
$\delta\Gamma/\Gamma \sim (H_{\ast}/2\pi\Phi)$~\cite{Dvali,Enqvist}. 
The fluctuating decay rate will give rise to a fluctuation in the 
reheat temperature of the Universe and this will be transmitted to 
the energy density of the thermalized plasma
\begin{equation}
\label{fact}
\frac{\delta \rho_{\gamma}}{\rho_{\gamma}}=-\frac{2}{3}\frac{\delta\Gamma}{
\Gamma}\equiv -\frac{2}{3}\frac{H_{\ast}}{2\pi\Phi}\,,
\end{equation} 
and the corresponding amplitude for the density perturbations will be given by
\begin{equation}
\label{fnl}
\delta_{\rm H}\sim \frac{2}{9}\frac{H_{\ast}}{2\pi\Phi}\,.
\end{equation}
The numerical factor $2/3$ appears in Eq.~(\ref{fact}) due to the fact
that the energy density of the inflaton oscillations goes as $a^{-3}$
while the radiation energy of the thermalized plasma falls as
$a^{-4}$. The factor $2/9$ in Eq.~(\ref{fnl}) is due to the fact that
the inflaton decays gives rise to the radiation dominated epoch.
Interestingly note that $\Phi$ can take large vevs near the end of
phase transition, as $\Phi\rightarrow (M_{p}/M_{\ast})\Phi_{0}$,
it is increasing probable to obtain the right amplitude for density
perturbations, which requires $\Phi\sim 10^{6}H_{\ast}$.

Another interesting possibility could be that the inflaton sector couples to 
the Standard Model sector via some Yukawa coupling 
$\kappa\phi HH+\phi(S/M_{p})qq+..$, where $H$ is the Standard Model Higgs,
$q$ Standard Model quarks, which has  
a non-renormalizable contribution of the form
\begin{equation}
\label{coupling}
\kappa=\kappa_{0} \left(1+\frac{\bf S}{M_{p}}+...\right)\,,
\end{equation}
where $S$ could be MSSM flat direction~\cite{Mazumdar1} or 
sneutrino~\cite{Mazumdar2}. In which case the inflaton coupling
$\kappa$ fluctuates by virtue of the fluctuation in $S$.

This gives rise to fluctuations in the energy density which is finally
imprinted in CMB. The fluctuations in the energy density of the 
relativistic species is given by \cite{Dvali,Enqvist,Mazumdar1,Marieke}
\begin{equation}
\label{coup2}
\frac{\delta \rho}{\rho} =-\frac{4}{3}\frac{\delta \kappa}{\kappa}=-\frac{4}{3}
\frac{\delta S}{S}\,.
\end{equation} 
Another useful way of imagining the coupling term $S$ in 
Eq.~(\ref{coupling}) as a fluctuating mass term for the 
inflaton. This completes our discussion on alternative ways of
generating adiabatic density perturbations.

%%%%%%%%%%%%%%%%%%%%%%%%%%%%%%%%%%%%%%%%%%%%%%%%%%%%%%%%%%%%%%%%%%%%%
\section{Conclusion}

In this paper we argue that it is possible to obtain extremely large
e-foldings of inflation at a fairly low scale. We exemplify this claim
by imposing extremely slow second order phase transition. This
realization can be implemented easily in hybrid inflation model where
usually inflation ends via phase transition. We also show that a large
Hubble induced inflaton mass is generically not a problem, because it
is possible to arrange the parameters such that enough inflation can
be obtained for galaxy formation during this phase transition. The
scale of phase transition can be brought down to the TeV scale, which
can ameliorate the cosmological moduli problem. We also discuss
various alternatives of generating adiabatic density perturbations,
especially we show that in hybrid model the orthogonal directions to
the inflaton can be responsible for generating adiabatic density
perturbations.

%%%%%%%%%%%%%%%%%%%%%%%%%%%%%%%%%%%%%%%%%%%%%%%%%%%%%%%%%%%%%%%%%%%%%%%%
\section*{Acknowledgments}
The author is thankful to Cliff Burgess, Jim Cline, Anne Green,
Abdel Lorenzana and Marieke Postma for discussion.
The author is a CITA national fellow.

%%%%%%%%%%%%%%%%%%%%%%%%%%%%%%%%%%%%%%%%%%%%%%%%%%%%%%%%%%%%%%%%%%%%%%%%  
 

\vskip20pt
\begin{references}
\bibitem{Linde}
See for example, A. D. Linde, {\it Particle Physics and Inflationary Cosmology},
New York, Harwood Academy (1990).

\bibitem{Ellis}
J. Ellis, J. E. Kim, and D. V. Nanopoulos, Phys. Lett. B {\bf 145}, 181 (1984)
%%CITATION = PHLTA,B145,181;%%

\bibitem{Maroto}
A.~L.~Maroto and A.~Mazumdar,
%``Production of spin 3/2 particles from vacuum fluctuations,''
Phys.\ Rev.\ Lett.\  {\bf 84}, 1655 (2000)
%%CITATION = HEP-PH 9904206;%%
R. Kallosh, L. Kofman, A. Linde, and A. Van Proyen,
Phys. Rev. D  {\bf 61}, 103503 (2000).
%%CITATION = HEP-TH 9907124;%%


\bibitem{Lyth}
D.~H.~Lyth and E.~D.~Stewart,
%``Cosmology with a TeV mass GUT Higgs,''
Phys.\ Rev.\ Lett.\  {\bf 75}, 201 (1995)
%%CITATION = HEP-PH 9502417;%%

\bibitem{Randall}
L.~Randall, M.~Soljacic and A.~H.~Guth,
%``Supernatural Inflation: Inflation from Supersymmetry with No (Very) Small Parameters,''
Nucl.\ Phys.\ B {\bf 472}, 377 (1996)
%%CITATION = HEP-PH 9512439;%%

\bibitem{garcia} 
J. Garc\'{\i}a-Bellido, A. D. Linde, and D. Wands, 
                 Phys. Rev. D {\bf 54}, 6040 (1996).

\bibitem{enqvist98}
K. Enqvist, and J. McDonald, Phys. Lett. B {\bf 425}, 309 (1998)
%%CITATION = HEP-PH 9711514;%%
A.~Kusenko and M.~E.~Shaposhnikov,
%``Supersymmetric Q-balls as dark matter,''
Phys.\ Lett.\ B {\bf 418}, 46 (1998)
%%CITATION = HEP-PH 9709492;%%

\bibitem{Reps}
K.~Enqvist and A.~Mazumdar,
%``Cosmological consequences of MSSM flat directions,''
Phys.\ Rept.\  {\bf 380}, 99 (2003)
%%CITATION = HEP-PH 0209244;%%
M.~Dine and A.~Kusenko,
%``The origin of the matter-antimatter asymmetry,''
arXiv:hep-ph/0303065.
%%CITATION = HEP-PH 0303065;%%


\bibitem{Linde1}
A. D. Linde, Phys. Lett. B {\bf 259}, 38 (1991); 
Phys. Rev. D {\bf 49}, 748 (1994).

\bibitem{new}
K. Freese, J. A. Frieman, A. V. Olinto, Phys. Rev. Lett. {\bf 65}, 3233 (1990).
.~C.~Adams, J.~R.~Bond, K.~Freese, J.~A.~Frieman and A.~V.~Olinto,
%``Natural inflation: Particle physics models, power law spectra for large scale structure, and constraints from COBE,''
Phys.\ Rev.\ D {\bf 47}, 426 (1993)
%%CITATION = HEP-PH 9207245;%%

\bibitem{angular}
G.~German, A.~Mazumdar and A.~Perez-Lorenzana,
%``Natural inflation from supergravity,''
Mod.\ Phys.\ Lett.\ A {\bf 17}, 1627 (2002)
%%CITATION = HEP-PH 0111371;%%





\bibitem{Copeland}
E.~J.~Copeland, A.~R.~Liddle, D.~H.~Lyth, E.~D.~Stewart and D.~Wands,
%``False vacuum inflation with Einstein gravity,''
Phys.\ Rev.\ D {\bf 49}, 6410 (1994)


\bibitem{softmass}
M.~Dine, W.~Fischler and D.~Nemeschansky,
%``Solution Of The Entropy Crisis Of Supersymmetric Theories,''
Phys.\ Lett.\ B {\bf 136}, 169 (1984);
%%CITATION = PHLTA,B136,169;
O.~Bertolami and G.~G.~Ross,
%``Inflation As A Cure For The Cosmological Problems Of Superstring Models With Intermediate Scale Breaking,''
Phys.\ Lett.\ B {\bf 183}, 163 (1987);
%%CITATION = PHLTA,B183,163;%%


\bibitem{dterm}
P.~Binetruy and G.~R.~Dvali,
%``D-term inflation,''
Phys.\ Lett.\ B {\bf 388}, 241 (1996),
%%CITATION = HEP-PH 9606342;%%
C.~F.~Kolda and J.~March-Russell,
%``Supersymmetric D-term inflation, reheating and Affleck-Dine  baryogenesis,''
Phys.\ Rev.\ D {\bf 60}, 023504 (1999),
%%CITATION = HEP-PH 9802358;%%


\bibitem{Green}
A.~M.~Green and A.~Mazumdar,
%``Dynamics of a large extra dimension inspired hybrid inflation model,''
Phys.\ Rev.\ D {\bf 65}, 105022 (2002)
%%CITATION = HEP-PH 0201209;%%

\bibitem{Abdel1} 
R. N. Mohapatra, A. P\'erez-Lorenzana and C. A. de S. Pires,
Phys. Rev. D {\bf  62}, 105030 (2000).A.~Mazumdar and A.~Perez-Lorenzana,
%``A dynamical stabilization of the radion potential,''
Phys.\ Lett.\ B {\bf 508}, 340 (2001)
%%CITATION = HEP-PH 0102174;%%
R.~Allahverdi, K.~Enqvist, A.~Mazumdar and A.~Perez-Lorenzana,
%``Baryogenesis in theories with large extra spatial dimensions,''
Nucl.\ Phys.\ B {\bf 618}, 277 (2001)
%%CITATION = HEP-PH 0108225;%%






\bibitem{fp} A. A. Starobinsky, in {\em Current topics in field theory, 
             quantum gravity and strings}, Lecture Notes in Physics,
             Eds. H. J. de Vega and N. Sanchez (Springer, Heidelberg 1986) 
             {\bf 206}, 107; A. S. Goncharov, A. D. Linde and V. F. Mukhanov,
             Int. J. Mod. Phys. {\bf A2} 561 (1987); A. D. Linde and 
             A. Mezhlumian, Phys. Lett. B {\bf 307}, 25 (1993);
             A. D. Linde, D. A. Linde and
             A. Mezhlumian, Phys. Rev. D {\bf 49} 1783 (1994).

\bibitem{Thomas}
M.~Dine, L.~Randall and S.~Thomas,
%``Baryogenesis from flat directions of the supersymmetric standard model,''
Nucl.\ Phys.\ B {\bf 458}, 291 (1996)
%%CITATION = HEP-PH 9507453;%%

\bibitem{msr} M. Sasaki and E. D. Stewart, Prog. Theor. Phys. {\bf 95}, 71 (1996)
\bibitem{Sloth}
K.~Enqvist and M.~S.~Sloth,
%``Adiabatic CMB perturbations in pre big bang string cosmology,''
Nucl.\ Phys.\ B {\bf 626}, 395 (2002)
%%CITATION = HEP-PH 0109214;%%

\bibitem{Lyth1}
D.~H.~Lyth and D.~Wands,
%``Generating the curvature perturbation without an inflaton,''
Phys.\ Lett.\ B {\bf 524}, 5 (2002),
%%CITATION = HEP-PH 0110002;%%
D.~H.~Lyth, C.~Ungarelli and D.~Wands,
%``The primordial density perturbation in the curvaton scenario,''
arXiv:astro-ph/0208055.
%%CITATION = ASTRO-PH 0208055;%%

\bibitem{Moroi}
T.~Moroi and T.~Takahashi,
%``Effects of cosmological moduli fields on cosmic microwave background,''
Phys.\ Lett.\ B {\bf 522}, 215 (2001)
[Erratum-ibid.\ B {\bf 539}, 303 (2002)]
%%CITATION = HEP-PH 0110096;%%

\bibitem{Mazumdar}
K.~Enqvist, S.~Kasuya and A.~Mazumdar,
%``Adiabatic density perturbations and matter generation from the MSSM,''
Phys.\ Rev.\ Lett.\  {\bf 90}, 091302 (2003)
%%CITATION = HEP-PH 0211147;%%
K.~Enqvist, A.~Jokinen, S.~Kasuya and A.~Mazumdar,
%``MSSM flat direction as a curvaton,''
arXiv:hep-ph/0303165.
%%CITATION = HEP-PH 0303165;%%

\bibitem{mcdonald}
N. Bartolo and A. R. Liddle, Phys. Rev. D {\bf 65}, 121301 (2003);
K. Dimopoulos and D. H. Lyth, hep-ph/0209180, M. Bastero-Gil, V. Di Clemente
and S. F. King, Phys. Rev. D {\bf 67} 103516 (2003); C. Gordon and A. Lewis,
Phys. Rev. D {\bf 67}, 123513 (2003);
K. Dimopoulos, D. H. Lyth, A. Notari and A. Riotto, JHEP, {\bf 0307} 053 (2003);K.~Hamaguchi, M.~Kawasaki, T.~Moroi and F.~Takahashi,
arXiv:hep-ph/0308174;
S.~Kasuya, M.~Kawasaki and F.~Takahashi,
arXiv:hep-ph/0305134; 
J. McDonald, Phys. Rev. D {\bf 68} 043503 (2003); J. McDonald, hep-ph/0310126.
K. Dimopoulos, G. Lazarides, D. H. Lyth and R. Ruiz de Austri,
hep-ph/0308015,M.~Axenides and K.~Dimopoulos, arXiv:hep-ph/0310194.




\bibitem{Lyth2}
D. H. Lyth, hep-th/0308110.

\bibitem{WMAP}
E.~Komatsu {\it et al.},
%``First Year Wilkinson Microwave Anisotropy Probe (WMAP) Observations: Tests of Gaussianity,''
arXiv:astro-ph/0302223.
%%CITATION = ASTRO-PH 0302223;%%

\bibitem{jaikumar}
P.~Jaikumar and A.~Mazumdar,
%``Post-inflationary thermalization and hadronization: QCD based approach,''
arXiv:hep-ph/0212265.
%%CITATION = HEP-PH 0212265;%%


\bibitem{Mazumdar1}
A.~Mazumdar and M.~Postma,
%``Evolution of primordial perturbations and a fluctuating decay rate,''
arXiv:astro-ph/0306509.
%%CITATION = ASTRO-PH 0306509;%%

\bibitem{Mazumdar2}
A.~Mazumdar,
%``A model for fluctuating inflaton coupling: (s)neutrino induced  adiabatic perturbations and non-thermal leptogenesis,''
arXiv:hep-ph/0306026.
%%CITATION = HEP-PH 0306026;%%
A.~Mazumdar,
%``CMB constraints on non-thermal leptogenesis,''
arXiv:hep-ph/0308020.
%%CITATION = HEP-PH 0308020;%%

\bibitem{Dvali}
G.~Dvali, A.~Gruzinov, and M.~Zaldarriaga, astro-ph/0303591;~~~
L.~Kofman, astro-ph/0303614; 
G.~Dvali, A.~Gruzinov and M.~Zaldarriaga, arXiv:astro-ph/0305548.


\bibitem{Enqvist}
K.~Enqvist, A.~Mazumdar and M.~Postma,
%``Challenges in generating density perturbations from a fluctuating  inflaton coupling,''
Phys.\ Rev.\ D {\bf 67}, 121303 (2003)
%%CITATION = ASTRO-PH 0304187;%%

\bibitem{Marieke}
M.~Postma and A.~Mazumdar,
%``Resonant decay of flat directions: Applications to curvaton scenarios,  Affleck-Dine baryogenesis, and leptogenesis from a sneutrinocondensate,''
arXiv:hep-ph/0304246.
%%CITATION = HEP-PH 0304246;%%


\end{references}
\end{document}